\def\gtsima{$\; \buildrel > \over \sim \;$}
\def\simgt{\lower.5ex\hbox{\gtsima}} 

\documentclass[
    ,final            
  ,numberedheadings 
  ]
  {aipproc}

\layoutstyle{6x9}
\usepackage{graphicx}

\begin{document}

\title{The Classification of Extragalactic X-ray Jets}


\author{D. E. Harris}{
  address={SAO, 60 Garden St., Cambridge, MA 02138, USA}
}

\author{F. Massaro}{
  address={SAO, 60 Garden St., Cambridge, MA 02138, USA}
}

\author{C. C. Cheung}{
  address={NRL/NRC, 4555 Overlook Ave SW, Washington, DC, 20375 USA}
}

\begin{abstract}
The overall classification of X-ray jets has clung to that prevalent
in the radio: FRI vs. FRII (including quasars).  Indeed, the common
perception is that X-ray emission from FRI's is synchrotron emission
whereas that from FRII's may be IC/CMB and/or synchrotron.  Now that
we have a sizable collection of sources with detected X-ray emission
from jets and hotspots, it seems that a more unbiased study of these
objects could yield additional insights on jets and their X-ray
emission.  The current contribution is a first step in the process of
analyzing all of the relevant parameters for each detected component
for the sources collected in the XJET website.  This initial effort
involves measuring the ratio of X-ray to radio fluxes and evaluating
correlations with other jet parameters.  For single zone synchrotron
X-ray emission, we anticipate that larger values of fx/fr should
correlate inversely with the average magnetic field strength (if the
acceleration process is limited by loss time equals acceleration
time).  Beamed IC/CMB X-rays should produce larger values of fx/fr for
smaller values of the angle between the jet direction and the line of
sight but will also be affected by the low frequency radio spectral
index.
\end{abstract}

\maketitle


\section{Defining the Problem}\label{sec:prob}

The paradigm for the classification of extragalactic jets is based on
the Fanaroff-Riley (\cite{fan:1974}) distinction between
powerful FRII radio galaxies (and radio quasars) on the one hand, and
low power FRI radio galaxies (and BL Lac objects) on the other.  The
(perhaps overly simplified) notion is that the jets of FRI sources are
lossy, often displaying a number of brightness enhancements
(a.k.a. "knots") whereas FRII jets in radio galaxies ("RG" hereafter)
are often difficult to detect and deposit the bulk of their energy at
large distances from the parent galaxy in the radio 'hotspots'.
Generally speaking, FRI jets often terminate within the confines of
their host galaxy (e.g. M87), or at least the well collimated section
of the jets are of similar size as their host (e.g. 3C~31).

Obviously this simple interpretation has a number of caveats; chief
amongst them is the effect of relativistic beaming which means that we
must make model-based assumptions to imagine what the jets of Cyg A
would look like if one of them was close to our line of sight
(l.o.s.).  In the unification hypothesis (\cite{urr:1995}) radio
quasars and FRII RG are distinguished solely on the basis of their
viewing angle.  If we were to view the jet of Cyg A close to the
l.o.s. would it be a knotty jet like that of 3C~273?

Our primary interest in this problem arises from the question of the
emission processes for the generation of X-rays from jets.  The idea
that kpc scale jets have substantial bulk Lorentz factors, $\Gamma
\simgt 5-10$ that would engender copious production of X-rays from
inverse Compton scattering of low energy relativistic electrons on
photons of the cosmic microwave background ("IC/CMB", \cite{tav:2000,
cel:2001, hk:2002}) came to be the accepted explanation for X-ray
emission from quasar and FRII RG jets.  For FRI jets, synchrotron
emission from very high energy electrons (i.e. electrons with Lorentz
factors $\gamma~\geq~10^7$) is thought to provide a consistent suite
of physical parameters and simple models were able to produce the
observed spectral energy distributions (SED).

Thus the classification of X-ray emitting jets tended to follow the
FRI - FRII separation, with the additional notion that FRI jets would
generally have the X-ray spectral index ($\alpha_x$, defined by flux
density, S$_{\nu}\propto\nu^{-\alpha}$) $>$1 whereas $\alpha_x$ for quasars
and FRII RG's would be $<$1.  These values of $\alpha_x$ are
consistent with expectations that at the high end of the electron
distribution (N($\gamma$)), we may expect steep spectra because we may
be observing close to an exponential cutoff and/or the initial
N($\gamma$) has been strongly affected at these energies by
synchrotron and IC losses which go as $\gamma^2$; and at the low
end of N($\gamma$), the emitted spectra (either synchrotron or IC)
would be flatter.

A number of doubts and counter examples have been raised about the
IC/CMB model of X-ray emission of quasar and FRII jets.  Amongst these
are: (a) the inability to independently demonstrate that
$\Gamma\geq$5 on kpc scales; (b) the assumption that one can
extrapolate N($\gamma$) from the radio regime ($\gamma\approx~10^4$)
down to $\gamma\approx$~100 with a power law of slope -p with
p=2$\times\alpha_{radio}$~+~1 (i.e. are there really enough low energy
electrons?); and (c) the evidence that the X-ray jet of 3C~273 comes
from a second synchrotron component rather than IC/CMB (\cite{uch:2006}).

These concerns have prompted us to attempt to devise a different
classification scheme for X-ray jets, hoping thereby to achieve a
better understanding of the role of IC and synchrotron emissions at
X-ray frequencies.  The present contribution reports on the first step of
this project: an investigation of the ratio of X-ray to radio flux for
X-ray detected knots and hotspots.  To do this, we have embarked on a
standard reduction scheme of all X-ray jets detected by the CXO
(Chandra X-ray Observatory).  The finding list for these sources comes
from the XJET compilation\footnote{http://hea-www.harvard.edu/XJET/}, see
Massaro et al., this volume.

\begin{figure}
\centerline{\includegraphics[height=8cm,width=0.55\linewidth,origin=c,angle=-90]{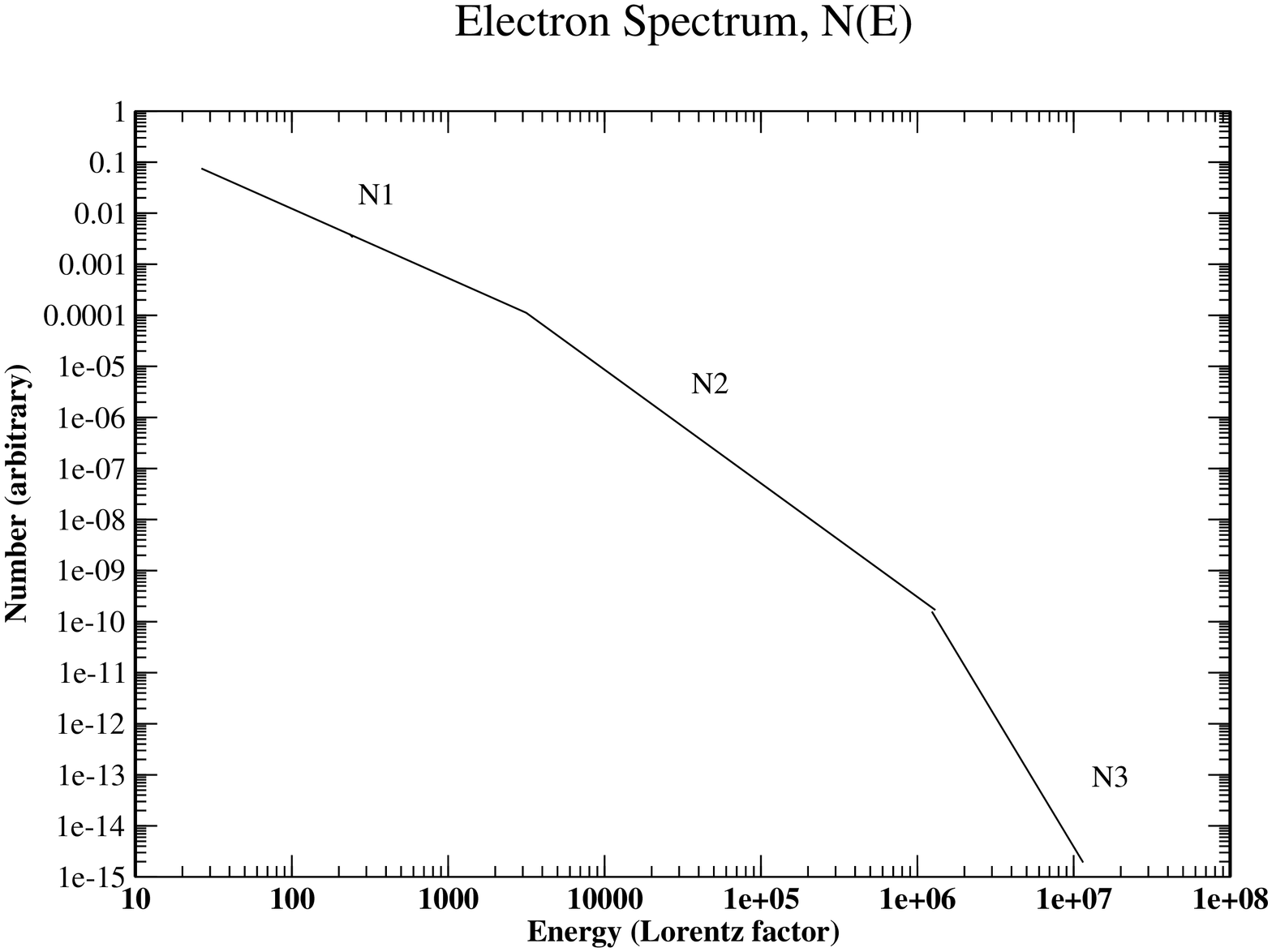}
\includegraphics[height=8cm,width=0.65\linewidth]{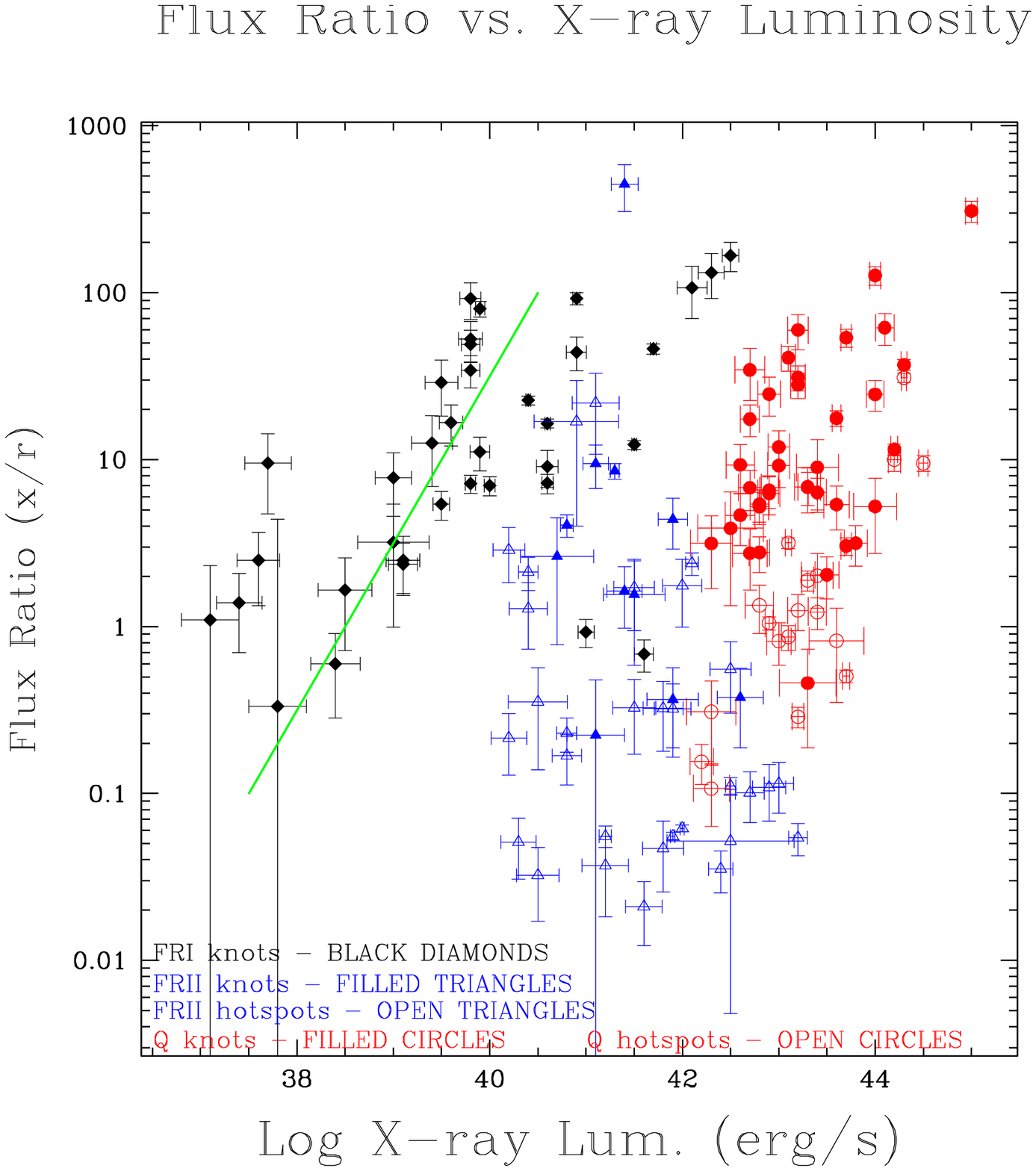}
}
\caption{Left: An idealized, broken power law electron spectrum.  N1 represents
the number of electrons contributing to IC/CMB X-rays; N2 the number
contributing to the radio emission; and N3, the number responsible for
synchrotron X-rays.  Right: A scatter plot of R versus X-ray
luminosity ($\Omega_{\Lambda}$=0.73, $\Omega_M$=0.23, and $H_{\rm 0} =
71$ km s$^{-1}$ Mpc$^{-1}$).  FRI knots are black diamonds; FRII knots
are blue filled triangles, and quasar knots are red filled circles.
Hotspots are open triangles (blue FRII) and circles (red Q).  The
green line is not a fit to anything; it has a slope of +1 (see
text).}\label{fig:lx}

\end{figure}

\section{Preliminary Results}

Our initial focus is an investigation of trends and correlations
involving the parameter R=$\frac{f_x}{f_r}$, the ratio of the X-ray
flux to the radio flux.  The former comes from our flux map photometry
(0.5-7 KeV, with no correction for absorption) and the latter is
calculated as $\nu~S_{\nu}$.  These radio fluxes come from VLA,
MERLIN, and ATCA maps at frequencies between 1.4 and 8 GHz.  

In order
to visualize how R may depend on various parameters, consider a
generalized electron spectrum consisting of 3 power laws
(fig.~\ref{fig:lx}, left panel): the lowest energy segment
characterized by an amplitude N$_1$, power law index p$_1$, covering
the range of $\gamma$'s from the tens to hundreds.  The next segment,
N$_2$ with a somewhat steeper spectrum would cover the energy range
responsible for the normal radio band via synchrotron emission.  The
third segment would cover the high end, $\gamma=10^6$ to 10$^8$.  The
number of electrons contributing to IC/CMB is meant to be represented
by N$_1$; the radio synchrotron by N$_2$, and the X-ray synchrotron by
N$_3$.  Obviously the relative values of the N's depends on the three
p's, the energies of any low and/or high energy cutoffs, and the
energies of the spectral breaks.

Considering the emission produced by this electron distribution,
the parameters relevant to the value of R can be isolated by:

\begin{equation}
R=\frac{X(ic) + X(sync)}{radio(sync)},
\end{equation}

\noindent
and for the case of beamed IC/CMB X-rays, of interest here,

\begin{equation}
R~\propto~\frac{f(\theta)~N_1~u(\nu) + N_3~B^2}{N_2~B^2}\propto~\frac{N_1}{N_2}\frac{f(\theta)\Gamma^2}{B^2} + \frac{N_3}{N_2}
\end{equation}

\noindent
where $f(\theta)$ is the extra beaming factor for IC/CMB
(\cite{hk:2002, mas:2009}); $\theta$ is the angle between the jet and
the line of sight; $u(\nu)$ is the photon energy density; $\Gamma$ is
the bulk Lorentz factor of the jet; and B is the average magnetic
field strength.
For FRI knots, the second term is thought to be the dominating contributor
so R is determined by the ratio of N's, which, in turn depends on $\alpha_2$,
$\alpha_3$, and the frequency of the break as well as the frequency of 
an exponential cutoff, if relevant.
If the X-ray emission of quasar knots (and hotspots) is primarily IC/CMB,
then the first term dominates and there is a significantly larger number of
(unknown) parameters.  In addition to the corresponding spectral parameters
of the synchrotron case for the ratio of N's, we have the unknown angle to the 
l.o.s., the unknown value of $\Gamma$, and the magnetic field strength.
A priori, it might seem that one should not expect to find any significant
correlation between R and other jet parameters.

We note that while the effect of the magnetic field strength is
explicit in the case of IC/CMB, it is also implicitly present for the
synchrotron case since a stronger field will usually produce a lower
frequency for the high energy cutoff as well as increasing the value
of $\alpha_3$.  Thus in general, we may expect stronger field regions
to have lower values of R. (than would otherwise be the case).

Because of space limitations, we show only one of our first set of
plots relating R to other jet properties.  The plot shown in the right
panel of fig.~\ref{fig:lx} (R as a function of X-ray luminosity) shows
a rather clean separation of FRI, FRII, and Q features.

If all features had the same one-to-one correspondence between X-ray
flux and radio flux, then all points would lie on a horizontal line.
The line in the figure has a slope of +1 and is shown to demonstrate
the locus of points which would describe knots that had the same radio
flux but different X-ray fluxes.  Obviously we do not expect either of
these conditions so we are surprised to find the majority of FRI
knots clustered about a particular +1 line.  Although the quasar
distribution follows the same trend, this behavior is not evident for
the FRII hotspots.

\section{Summary}

From our preliminary analysis using the ratio of X-ray to radio
fluxes, we have found that hotspots tend to have a lower ratio than
knots.  Features classified as 'quasar hotspots' generally have
smaller values of R than quasar knots, and most FRII hotspots have
smaller R values than FRI and quasar knots.  Most FRI and quasar knots
have R values between 1 and 100.  We have failed to find any
significant distinction between FRI and quasar knots except for the
obvious difference in apparent x-ray luminosity.


\begin{theacknowledgments}

We are indebted to a number of colleagues who have generously donated
radio maps for our use, and for public access via the XJET webpage.
Other radio maps were downloaded from the NRAO VLA Archive Survey.
The work at SAO was partially supported by NASA grant AR6-7013X. FM
acknowledges the Foundation BLANCEFLOR Boncompagni-Ludovisi, n'ee
Bildt for the grant awarded him in 2009.

\end{theacknowledgments}


\small
\begin{thebibliography}{9}

\bibitem{cel:2001}
A. Celotti, G. Ghisellini, \& M. Chiaberge,  \emph{MNRAS} \textbf{321}, L1--5 (2001)

\bibitem{fan:1974}
B.~L. Fanaroff \& J.~M. Riley, J. M., \emph{MNRAS} \textbf{167}, 31p--36p (1974)

\bibitem{hk:2002}
D.~E. Harris and H. Krawczynski, \emph{ApJ} \textbf{565}, 244--255 (2002)


\bibitem{mas:2009}
F. Massaro, D.~E. Harris, M. Chiaberge, P. Grandi, F.~D. Macchetto,
S.~A. Baum, C.~P. O'Dea, A. Capetti, \emph{ApJ} \textbf{696}, 980--985
(2009)

\bibitem{tav:2000}
F. Tavecchio, L. Maraschi, R.~M. Sambruna, \& C.~M. Urry, \emph{ApJL} \textbf{544},
L23--26 (2000)

\bibitem{uch:2006} Y. Uchiyama et al.,  \emph{ApJ} \textbf{648}, 910-921  (2006)

\bibitem{urr:1995}
C.~M. Urry \& P. Padovani, \emph{PASP} \textbf{107}, 803--845 (1995)

\end{thebibliography}
\end{document}